# 重力及重力梯度快速正演

曹书锦

1. 湖南科技大学 页岩气资源利用湖南重点实验室，湘潭 411201；

2. 湖南科技大学 资源环境与安全工程学院，湘潭 411201；

**摘　要**：重力场正演计算效率制约着大尺度重力资料的反演解释，所以研究如何快速且精确地实现重力场正演计算方法对野外实测重力资料解释非常必要。为避免存储反演核矩阵，在等效几何构架的基础上，分析其依赖重力场对称性的原因；进一步引入平移等效性的扩充技术，从而移除等效几何构架中对位场对称性的依赖，以构建适合水平观测面的快速正演方法。数值试验表明该方法计算精度高、计算效率好和内存消耗小。

**关键字**：等效几何构架；重力梯度张量；平移等效性；快速正演

## Rapid forward of gravity and tensor gravity data

CAO Shu-jin

1. Hunan Provincial Key Laboratory of Shale Gas Resource Utilization, Hunan University of Science and Technology, Xiangtan 411201, China;

2. School of Resource Environment and Safety Engineering, Hunan University of Science and Technology, Xiangtan 411201, China

**Abstract**：The large-scale three-dimensional inversion of surface gravity / tensor gravity data is a very challenging numerical and practical problem, which is a highly physical memory usage, time-consuming computation and high precision for large-scale gravity models. Equivalent geometric architecture was introduced to avoid to calculate and to save sensitivity matrix. A new equivalent geometric architecture was not rely on symmetrical characteristic of gravity field by added three virtual grids into geophysical model. A new fast forward method was proposed based translation equivalent technique based toeplitz matrix, which a toeplitz matrix carried out fast matrix-vector multiplication by using the fast Fourier transform. Numerical experiments show that the method proposed in this paper is require little memory, high efficiency and high precision.

**Keywords**：equivalent geometric architecture; full tensor gravity; translation equivalent technique; rapid forward

# 1 绪 论

重力/重力梯度正演计算效率制约着大尺度重力资料反演解释(陈召曦等，2012；吴文鹂等, 2009；姚长利等，2003)。目前构建正演算子主要有如下两种途径：一、基于积分方程，以有解析解的简单规则体对地质体进行离散，基于位场可叠加原理，通过叠加各简单规则体的异常响应获得重力场，如面元法和点元法等(曹书锦等，2010)；二、基于偏微分方程，利用如有限差分法(Farquharson and Mosher，2009；Camacho et al., 2011)、有限元法(徐世浙，2005)和有限体积法(Jahandari and Farquharson，2013)等数值模拟技术对重力场泊松方程进行离散，通过解大型稀疏方程组获得重力场。相比于前者，由于泊松方程右端项不为常量，且右端项随着密度不为零的物性网格数目的增多而变得复杂，导致解大型稀疏方程组的耗时急剧增加。因而重力/重力梯度数据反演解释多采用点元法，但当模型规模较大、网格剖分数较多或观测点数较多时，正演计算存在计算量急剧增加和内存消耗过大等问题。

为此，很多学者自如下两个方面开展重力/重力梯度快速正演研究。一方面，在不改变计算量的前提下，开展重力/重力梯度场正演并行计算。如陈召曦等(陈召曦等，2012；Moorkamp et al., 2010)采用GPU并行计算技术实现了重力和重力梯度分量快速正演计算。由于不储存核矩阵，这导致了反演计算效率相对较低。另一方面，在不更换计算设备的前提下，开展核矩阵压缩技术研究。如Li等(Li and Oldenburg，2003)通过小波变换来降低核矩阵内存消耗，Mukherjee et al.(2004)采用质点球的解析解替代直立六面体的解析解以加快正演计算。但这两种方法均在一定程度上损失了信号精度(陈召曦等，2012；吴文鹂等，2009)。姚长利等(姚长利等，2003)指出若将观测点与物性网格一一对应，仅需正演计算一次最小观测点对所对应的物性网格的核函数，即等效几何构架技术，但其依赖于重力场的对称性。

由于等效几何构架依赖重力场的对称性，这导致了其难于推广至重力梯度正演计算。为此，引入基于平移等效性的扩充技术，避免了等效几何构架对重力场及重力梯度场的对称性产生依赖。通过单条测线与单列物性网格所构建的核矩阵分析，证明该矩阵为典型的Toeplitz矩阵，进一步证明所有观测点对应于单层物性网格的核矩阵为Block Toeplitz Toeplitz Block（BTTB）矩阵。由于Toeplitz矩阵可通过快速傅里叶变换（FFT）能快速实现矩阵与向量的乘法，在此基础上，本研究构建适合水平观测面下的快速正演方法。试验结果表明，该方法的计算结果与常规算法的计算结果一致，可极大地减少内存消耗和大大提高计算效率。

# 2 重力梯度张量正演计算技术

## 2.1 传统重力梯度张量正演解析解

重力测量只观测到重力位的铅垂一次导数$\Delta g$，而重力梯度测量可以得到重力位的一次导数$g_x$、$g_y$和$g_z$在笛卡尔坐标系$X$、$Y$和$Z$等三轴向上的变化率，即重力位的二阶导数(Mickus and Hinojosa, 2001; Mukherjee et al., 2004)，则重力梯度可写为：

$$\boldsymbol{T} = \begin{bmatrix} [g_{xx}] & [g_{xy}] & [g_{xz}] \\ g_{yx} & [g_{yy}] & [g_{yz}] \\ g_{zx} & g_{zy} & g_{zz} \end{bmatrix} \quad (1)$$

由于重力场为无旋场和无散场(Telford and Sheriff, 1990; Blakely, 1996)，根据场论可知：$g_{xy}=g_{yx}$、$g_{xz}=g_{zx}$、$g_{yz}=g_{zy}$ 和 $g_{zz}=-(g_{xx}+g_{yy})$，所以 $\boldsymbol{T}$ 仅有5个独立分量，即式(1)中方框内的元素。

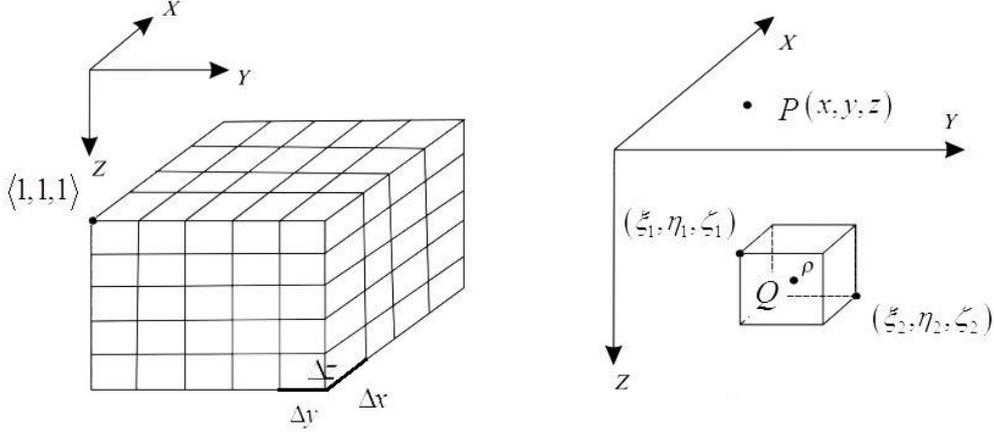

(a) 地下剖分单元示意图, (b) 测量示意图

图 1 地下剖分模型示意图

Fig. 1 Interpretation models of the subsurface. (a) discrete schematic diagram of interpretation models, (b) Gravitational attraction at the survey point P due to single prism

在基于点元法的传统重力/重力梯度场物性反演方案中，仅涉及规则长方体网格单元的正演计算。如图 1所示，利用三边长为$\Delta x$、$\Delta y$和$\Delta z$的长方体将地质体离散，其沿笛卡尔坐标系三轴向剖分数分别为$n_x$、$n_y$和$n_z$，观测点个数为$N_d = n_x \times n_y$，物性网格个数为 $N_m = n_x \times n_y \times n_z$，并以$\langle p,q \rangle$（对于等维反演而言(张永明和张贵宾， 2009)为$\langle p,q,t \rangle$）和$\langle l,m,n \rangle$分别对任一观测点$P$及任一物性网格$Q$按其所处空间维度进行编号。限于篇幅，仅以$\Delta g$为例，密度值为$\rho_{(\xi,\eta,\zeta)}$的物性网格$Q$在观测点$P$处的异常响应的计算公式为(Li and Chouteau，1998)：

$$\Delta g = -\upsilon \rho_{(\xi,\eta,\zeta)} \sum_{i=1}^{2}\sum_{j=1}^{2}\sum_{k=1}^{2} \mu_{ijk} \left[ x_i \ln\left( \frac{y_j}{\sqrt{x_i^2+z_k^2}} + \sqrt{1+\frac{y_j^2}{x_i^2+z_k^2}} \right) \right.$$
$$\left. + y_j \log\left( \frac{x_i}{\sqrt{y_j^2+z_k^2}} + \sqrt{1+\frac{x_i^2}{y_j^2+z_k^2}} \right) - z_k \arctan\frac{x_i y_j}{z_k r_{ijk}} \right] \quad (2)$$

其中，$(\xi_1,\eta_1,\zeta_1)$和$(\xi_2,\eta_2,\zeta_2)$分别为长方体两对角顶点坐标；$x_i = x-\xi_i$，$y_j = y-\eta_j$，$z_k = z-\zeta_k$；$r_{ijk} = \sqrt{x_i^2+y_j^2+z_k^2}$，$\mu_{ijk} = (-1)^{i+j+k}$；$\upsilon = 6.672\times10^{-11}\,\text{N}\cdot\text{m}^2/\text{kg}^2$为引力常数。

基于位场可叠加原理，对于所有物性网格在第$i$个观测点处的异常响应，可记为：

$$d_i = \sum_{j=1}^{N_m} G_{ij}\rho_j \tag{3}$$

其中，$d_i$为第$i$个观测点处的重力/重力梯度异常值，$G_{ij}$为第$j$个物性网格在第$i$个观测点处的核函数，$\rho_j$为第$j$个物性网格的密度值。

写成矩阵形式：

$$\boldsymbol{d}=\boldsymbol{GM} \tag{4}$$

式中：**d**为数据向量，**G**为核矩阵，**M**为模型向量。重力或重力梯度正演问题是已知**G**和**M**，求解**d**；而相应的反演问题则是已知**G**和**d**，求解**M**。因而，点元法中为存储核矩阵**G**而需要存储的核函数的个数为

$$N_G = N_M \times N_d = (n_x \times n_y) \cdot (n_x \times n_y \times n_z) \tag{5}$$

这里，乘号·定义为观测点数与网格个数的乘法。

## 2.2 重力梯度张量快速正演

重力/重力梯度数据的三维正演：对于常规勘探方法而言，观测数据**d**是二维的，物性网格**M**是三维的；对于等维反演而言，观测数据**d**是三维的，物性网格**M**也是三维的。以图 2 中两类位场勘探方法为例，两者的观测点个数大体一致。由式(4)可知$N_G = (1024 \times 1024 \times 100) \times (1024 \times 1024 \times 100) \approx 1.1 \times 10^{16}$。无论是采用双精度还是单精度存储，都远超现有个人计算机的物理内存容量。针对这一问题，姚长利等(2003)指出在式(2)中仅$\rho_{(\xi,\eta,\zeta)}$随着反演迭代而变化，而剩余部分为几何构架。现将式(2)重写为几何构架$K_{(\xi,\eta,\zeta)}^{(x,y,z)}$和密度$\rho_{(\xi,\eta,\zeta)}$两部分：

$$\Delta g = K_{(\xi,\eta,\zeta)}^{(x,y,z)} \rho_{(\xi,\eta,\zeta)} \tag{6}$$

其中

$$K_{(\xi,\eta,\zeta)}^{(x,y,z)} = -\upsilon \sum_{i=1}^{2}\sum_{j=1}^{2}\sum_{k=1}^{2} \mu_{ijk} \left[ x_i \ln\left(\frac{y_j}{\sqrt{x_i^2+z_k^2}} + \sqrt{1+\frac{y_j^2}{x_i^2+z_k^2}}\right) \right.$$
$$\left. + y_j \log\left(\frac{x_i}{\sqrt{y_j^2+z_k^2}} + \sqrt{1+\frac{x_i^2}{y_j^2+z_k^2}}\right) - z_k \arctan\frac{x_i y_j}{z_k r_{ijk}} \right]$$

式中：几何构架$K_{(\xi,\eta,\zeta)}^{(x,y,z)}$的上标$(x, y, z)$对应于任一观测点$P$，下标$(\xi,\eta,\zeta)$对应于任一物性网格$Q$的中心点坐标，$K_{(\xi,\eta,\zeta)}^{(x,y,z)}$的物理含义为物性网格$Q$在观测点$P$处的核函数。

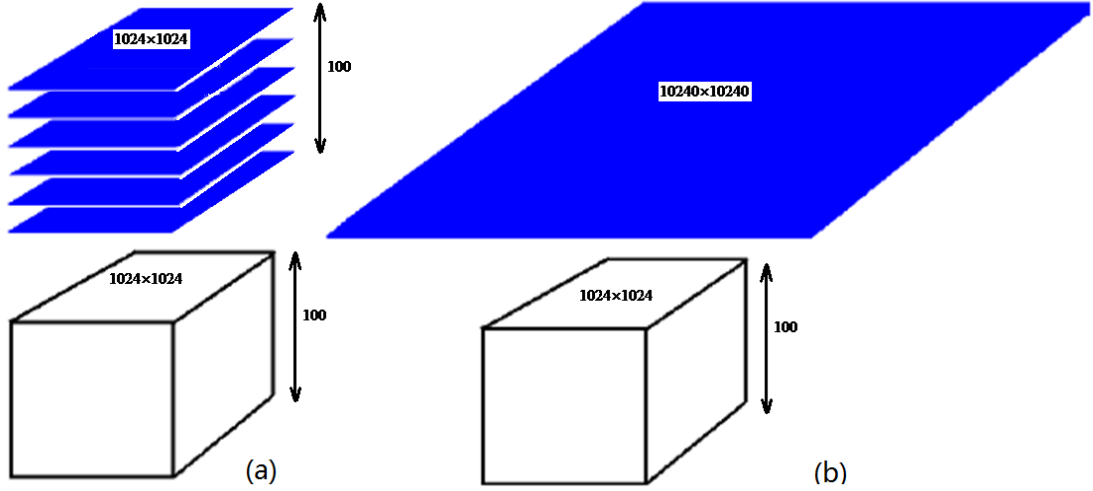

图 2 两类位场勘探观测点布设方案简要示意图。(a) 等维反演，(b) 常规反演

Fig. 2 Two cases of potential field exploration. (a) equivalent dimensions inversion, (b) normal inversion

为更清晰地标示第$i$个观测点$P$与第$j$个物性网格$Q$之间关系，将几何构架$K_{(\xi,\eta,\zeta)}^{(x,y,z)}$以观测点序号$\langle p,q,t \rangle$和物性网格编号$\langle l,m,n \rangle$来重新表述$\overline{K}_{\langle l,m,n \rangle}^{\langle p,q,t \rangle}$，即对应于核函数$K^{i,j}$：

$$K^{i,j} \equiv \overline{K}_{\langle l,m,n \rangle}^{\langle p,q,t \rangle} \equiv K_{(\xi,\eta,\zeta)}^{(x,y,z)}$$

式中，$i = p+(q-1)n_x + (q-1)n_x \times n_y$，$j = l+(m-1)n_x + (n-1)n_x \times n_y$；$1 \leq l \leq n_x$，$1 \leq m \leq n_x$，$1 \leq p \leq n_y$，$1 \leq q \leq n_x$，$1 \leq n \leq n_z$，$t \geq 1$。

针对$K_{(\xi,\eta,\zeta)}^{(x,y,z)}$在反演过程中没有变化这一特点，姚长利等(2003)指出几何构架$K_{(\xi,\eta,\zeta)}^{(x,y,z)}$具有平移等效性、位场对称性和互换等效性。所谓的互换等效性即指当观测点$P$与物性网格$Q$两者空间位置互换时，核函数不发生变化，可表述为(姚长利等，2003)

$$\overline{K}_{\langle l,m,n \rangle}^{\langle p,q,t \rangle} = \overline{K}_{\langle p,q,t \rangle}^{\langle l,m,n \rangle}$$

在笛卡尔坐标系中，任意关于$X$轴对称的两点$(-x,y,z)$和$(x,y,z)$处的核函数相等，即位场对称性，可表述为(姚长利等，2003)

$$\overline{K}_{\langle l,m,n \rangle}^{\langle p,q,t \rangle} = \overline{K}_{\langle l,m,n \rangle}^{\langle |p|,q,t \rangle}$$

同样地，当关于$Y$轴对称时，

$$\overline{K}_{\langle l,m,n \rangle}^{\langle p,q,t \rangle} = \overline{K}_{\langle l,m,n \rangle}^{\langle p,|q|,t \rangle}$$

所谓的平移等效性即指当观测点$P$与物性网格$Q$的相对空间位置没有变化时，无论观测点$P$或物性网格$Q$移动到何处，其核函数值始终保持不变，可表述为(姚长利等，2003)

$$\begin{cases} \overline{K}_{\langle l,m,n \rangle}^{\langle p,q,t \rangle} = \overline{K}_{\langle l+1,m,n \rangle}^{\langle p+1,q,t \rangle} = \cdots = \overline{K}_{\langle l+\Delta p,m,n \rangle}^{\langle p+\Delta p,q,t \rangle} \\ \overline{K}_{\langle l,m,n \rangle}^{\langle p,q,t \rangle} = \overline{K}_{\langle l,m+1,n \rangle}^{\langle p,q+1,t \rangle} = \cdots = \overline{K}_{\langle l,m+\Delta q,n \rangle}^{\langle p,q+\Delta q,t \rangle} \\ \overline{K}_{\langle l+\Delta p,m+1,n \rangle}^{\langle p+\Delta p,q+1,t \rangle} = \overline{K}_{\langle l+\Delta p,m+2,n \rangle}^{\langle p+\Delta p,q+2,t \rangle} = \cdots = \overline{K}_{\langle l+\Delta p,m+\Delta q,n \rangle}^{\langle p+\Delta p,q+\Delta q,t \rangle} \\ \cdots \end{cases} \quad (7)$$

式中，$1 \leq \Delta p+p \leq n_x$，$1 \leq \Delta p+l \leq n_x$，$1 \leq \Delta q+q \leq n_y$，$1 \leq \Delta q+m \leq n_y$。

### 2.2.1 位场对称性分析

根据式(7)，将所有观测点偏移到序号为$\langle 1,1,1\rangle$的等效几何构架计算原点(如图 1所示，即为观测网的坐标值最小点)，则$\overline{K}_{\langle l,m,n\rangle}^{\langle p,q,t\rangle}$可以重写为：

$$\overline{K}_{\langle l,m,n\rangle}^{\langle p,q,t\rangle} = \overline{K}_{\langle |l-p|+1,|m-q|+1,n-t+1\rangle}^{\langle 1,1,1\rangle} \tag{8}$$

式(8)被称之为等效几何构架计算公式(姚长利等，2003)。但由于式(8)使用了绝对值标记$|\cdot|$，这使得正演计算依赖于位场的对称性，即$|l-p|$依赖于位场关于$Y$轴对称和$|m-q|$依赖于位场关于$X$轴对称。由式(8)可知，在计算重力场$g_z$时，仅需存储等效几何构架计算原点关于所有物性网格的核函数。因而，等效几何构架需要存储核函数个数为

$$N_G = N_M \times N_d = (1) \cdot (n_x \times n_y \times n_z) \tag{9}$$

### 2.2.2 扩充平移等效性

如图 3所示，在等效几何构架计算策略下，首先将原有观测网$S$（黑色线框）内任一观测点$\langle p,q,t\rangle$沿着红色路径偏移至计算原点$\langle 1,1,1\rangle$，新观测网$S'$（蓝色线框）由1、2、3和4块构成，其中处于第1、2和3块位置物性网格计算于的异常响应需要依赖于位场的对称性，即需要将第1、2和3块绕$X$轴、$Y$轴和计算原点翻转至$S$内的第1'、2'和3'块进行计算。

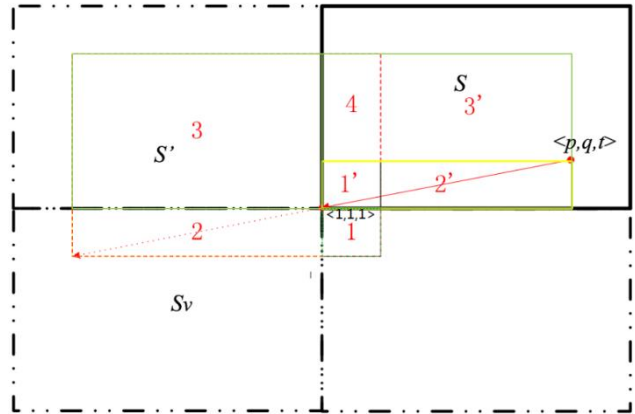

图 3 虚拟物性网格设置示意图

Fig. 3 Schematic diagram of virtual grid

为不依赖于位场的对称性，在关于$X$轴、$Y$轴和计算原点对称的方位添加三个与原有地下物性区域一样大小的虚拟网格(两点划线)，从而三个虚拟网格与原有网格$S$构建虚拟物性网格$Sv$。在$Sv$上移动$S$，使得任意观测点$\langle p,q,t\rangle$与计算原点$\langle 1,1,1\rangle$重合，从而获得新观测网$S'$。这使得$S'$在虚拟物性网格上$Sv$仅依赖于平移等效性计算，而不依赖于位场对称性。基于平移等效性的等效几何构架计算公式为：

$$\overline{K}_{\langle l,m,n\rangle}^{\langle p,q,t\rangle} = \overline{K}_{\langle \tilde{l},\tilde{m},n-t+1\rangle}^{\langle 1,1,1\rangle} \tag{10}$$

式中：$\tilde{l} = l-p+1$，$\tilde{m} = m-q+1$，$-n_x+1 \leqslant \tilde{l} \leqslant n_x$，$-n_y+1 \leqslant \tilde{m} \leqslant n_y$。在重力\重力梯度正演计算中$g_z$时，仅需存储等效几何构架计算原点关于$Sv$的核函数。因而，需要存储核函数个数为：

$$N_G = N_M \times N_d = (1) \cdot (2n_x \times 2n_y \times 2n_z) \tag{11}$$

### 2.2.3 快速算法实现

根据式(6)，将式(4)按照式(3)逐一观测点写为矩阵形式

$$\begin{pmatrix} K^{1,1} & K^{1,2} & \cdots & K^{1,j} & \cdots & K^{1,N_m} \\ K^{2,1} & K^{2,2} & \cdots & K^{2,j} & \cdots & K^{2,N_m} \\ \vdots & \vdots & \ddots & \vdots & \ddots & \vdots \\ K^{i,1} & K^{i,2} & \cdots & K^{i,j} & \cdots & K^{i,N_m} \\ \vdots & \vdots & \ddots & \vdots & \ddots & \vdots \\ K^{N_d,1} & K^{N_d,2} & \cdots & K^{N_d,j} & \cdots & K^{N_d,N_m} \end{pmatrix} \begin{pmatrix} \rho_1 \\ \rho_2 \\ \vdots \\ \rho_j \\ \vdots \\ \rho_{N_m} \end{pmatrix} = \begin{pmatrix} d_1 \\ d_2 \\ \vdots \\ d_i \\ \vdots \\ d_{N_d} \end{pmatrix} \tag{12}$$

其中，$\rho_j$ 为第 $j$ 个物性网格的密度值。

为便于显示，如图 4所示，即在图 2中的右图中标注第 $q$ 条测线（黑线）和第 $m$ 列网格（红线）。

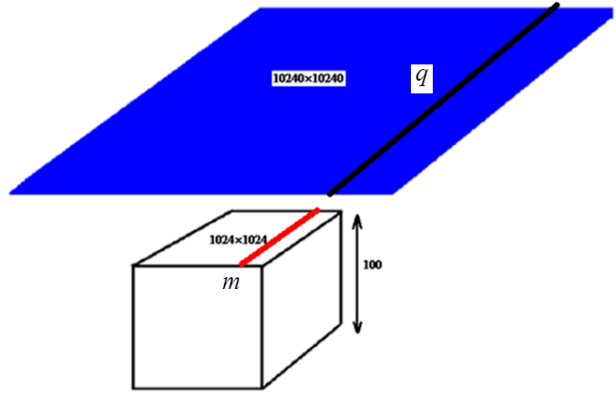

图 4 位场勘探测线示意图

Fig. 4 sketch of prospecting line

以图 4中的第 $q$ 条测线及所对应的第 $m$ 列网格的观测方案为例，仿写式(12)

$$\begin{pmatrix} K^{i'+1,j'+1} & K^{i'+1,j'+2} & \cdots & K^{i'+1,j'+\Delta m} & \cdots & K^{i'+1,j'+n_y} \\ K^{i'+2,j'+1} & K^{i'+2,j'+2} & \cdots & K^{i'+2,j'+\Delta m} & \cdots & K^{i'+2,j'+n_y} \\ \vdots & \vdots & \ddots & \vdots & \ddots & \vdots \\ K^{i'+\Delta q,j'+1} & K^{i'+\Delta q,j'+2} & \cdots & K^{i'+\Delta q,j'+\Delta m} & \cdots & K^{i'+\Delta q,j'+n_y} \\ \vdots & \vdots & \ddots & \vdots & \ddots & \vdots \\ K^{i'+n_y,j'+1} & K^{i'+n_y,j'+2} & \cdots & K^{i'+n_y,j'+\Delta m} & \cdots & K^{i'+n_y,j'+n_y} \end{pmatrix} \begin{pmatrix} \rho_{j'+1} \\ \rho_{j'+2} \\ \vdots \\ \rho_{j'+\Delta m} \\ \vdots \\ \rho_{j'+n_y} \end{pmatrix} = \begin{pmatrix} d_{i'+1} \\ d_{i'+1} \\ \vdots \\ d_{i'+\Delta q} \\ \vdots \\ d_{i'+n_y} \end{pmatrix} \tag{13}$$

或

$$\boldsymbol{G}_{q,m} \boldsymbol{\rho}_m = \boldsymbol{d}_q \tag{14}$$

式中：$\boldsymbol{G}_{q,m}$ 为第 $q$ 条测线关于第 $m$ 列物性网格的核矩阵，$\boldsymbol{\rho}_m$ 为第 $m$ 列物性网格的密度值，$\boldsymbol{d}_q$ 为第 $m$ 列物性网格在第 $q$ 条测线上引起的异常值，$i' = (q-1) \times n_x + (t-1) \times n_x \times n_y$，$j' = (m-1) \times n_x + (n-1) \times n_x \times n_y$。

基于式(7)，以任意一 $\Delta q$ 和 $\Delta m$ 步长分别移动观测点 $P$ 和物性网格 $Q$ 都能在式(13)左侧矩

阵中找到各元素相等的对角线，如当$\Delta q=1$且$\Delta m=1$时：

$$K^{i+1,j+1} = K^{i+2,j+2} = \cdots = K^{i+\Delta q,j+\Delta q} = K^{i+\Delta m,j+\Delta m} = K^{i+n_y,j+n_y} \tag{15}$$

或当$\Delta q=1$且$\Delta m=2$时：

$$K^{i+1,j+2} = K^{i+2,j+3} = \cdots = K^{i+\Delta q,j+\Delta q+1} = K^{i+\Delta m,j+\Delta m+1} = K^{i+n_y-1,j+n_y} \tag{16}$$

按照相类似的方法，可构建其它对角线。因而，$G_{q,m}$可以由各对角线第一个元素表达，可写为：

$$G_{q,m} = \begin{pmatrix} a_0 & a_1 & \cdots & \cdots & a_{n_y-1} \\ a_{-1} & a_0 & a_1 & & \vdots \\ \vdots & a_{-1} & a_0 & \ddots & \vdots \\ \vdots & & \ddots & \ddots & a_1 \\ a_{1-n_y} & \cdots & \cdots & a_{-1} & a_0 \end{pmatrix} \tag{17}$$

式中，$[a_0, a_{-1}, a_{-2}, \cdots, a_{1-n_y}]$对应于式(13)中矩阵$G_{q,m}$的第一列，$[a_0, a_1, a_2, \cdots, a_{n_y-1}]$对应于式(13)中矩阵$G_{q,m}$的第一行，即通过$2n_y-1$个元素可以表达$G_{q,m}$。

由Toeplitz矩阵定义可知(Chan and Jin, 2007)，$G_{q,m}$为Toeplitz矩阵(见图 5)。根据Toeplitz矩阵与傅里叶级数关系，$G_{q,m}\rho_m$的乘积可写为：

$$G_{q,m}\rho_m = \mathcal{F}\left(F(a^{\mathrm{T}}) \cdot \mathcal{F}\left(\begin{bmatrix} \rho_m \\ 0 \end{bmatrix}\right)\right) \tag{18}$$

式中，$\mathcal{F}$与$F$为傅里叶变换对；T为矩阵转置标记；作为替代核矩阵而需要存储的部分$a$为：

$$a = \left\{[a_0, a_1, a_2, \cdots, a_{n_y-1}] \quad 0 \quad \mathcal{F}\left([a_{-1}, a_{-2}, \cdots, a_{1-n_y}]\right)\right\} \tag{19}$$

类似于式(17)，可以构建其它测线与各列网格的关系，并将第$t$层观测点及所对应的第$n$层物性网格的核矩阵写为

$$G_{t,n} = \begin{pmatrix} G_{1,1} & G_{1,2} & \cdots & \cdots & G_{1,n_x} \\ G_{2,1} & G_{2,2} & G_{2,3} & & \vdots \\ \vdots & G_{3,2} & G_{3,3} & \ddots & \vdots \\ \vdots & & \ddots & \ddots & G_{n_x-1,n_x} \\ G_{n_x,1} & \cdots & \cdots & G_{n_x,n_x-1} & G_{n_x,n_x} \end{pmatrix}$$

根据平移等效性，类似于式(15)~(16)的推导过程，可以导出$G_{t,n}$中各对角线上的子矩阵相等。因而$G_{t,n}$可以按照式(17)写为

$$G_{t,n} = \begin{pmatrix} b_0 & b_1 & \cdots & \cdots & b_{n_y-1} \\ b_{-1} & b_0 & b_1 & & \vdots \\ \vdots & b_{-1} & b_0 & \ddots & \vdots \\ \vdots & & \ddots & \ddots & b_1 \\ b_{1-n_y} & \cdots & \cdots & b_{-1} & b_0 \end{pmatrix}$$

其中，$[b_0, b_{-1}, b_{-2}, \cdots, b_{1-n_y}]$ 对应于式(13)中矩阵 $G_{t,n}$ 第一列，$[b_0, b_1, b_2, \cdots, b_{n_y-1}]$ 对应于式(13)中矩阵 $G_{t,n}$ 的第一行，即通过 $2n_y-1$ 个子矩阵可以表达 $G_{t,n}$。那么内存占用比率至多为原有的：

$$Ratio_{G_{t,n}} = \frac{(2n_x-1)}{n_x^2} \frac{(2n_y-1)}{n_y^2} \tag{20}$$

由BTTB矩阵定义可知(Chan and Jin, 2007)，$G_{t,n}$ 为BTTB矩阵(见图 5)，$G_{t,n}$ 与相应物性网格密度值的乘积与式(18)类似，此处不再赘述。由于重力位一阶导数和重力梯度均具有平移等效性，因而 $g_x$、$g_y$ 及重力梯度各分量的核矩阵均为BTTB矩阵。

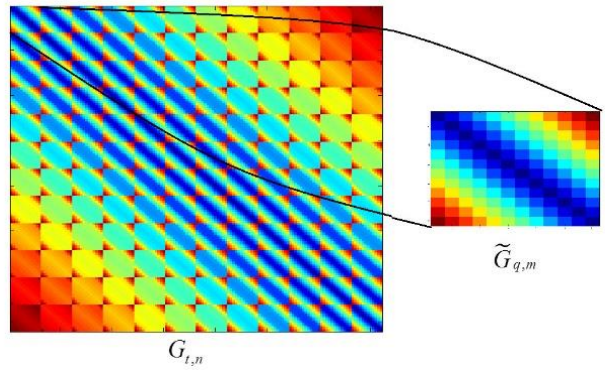

图 5 单层观测点所对应于单层物性网格的核矩阵示意图

Fig. 5 Schematic diagram of $g_{xx}$ with single layer mesh for all survey point

对于常规勘探观测方案而言，一般地，由单层观测点(即 $t = 1$ 时)和多层物性网格构成核矩阵 $G$，即核矩阵 $G$ 可以由多个 $G_{t,n}$ 子矩阵组成的向量构成：

$$G = \begin{bmatrix} G_{1,1} & G_{1,2} & \cdots & G_{1,n} \end{bmatrix} \tag{21}$$

由(19)~(21)式可知，替代存储核矩阵 $G$ 而需要存储元素个数为

$$N_G = N_m \times N_d = (1) \cdot ((2n_x-1) \times (2n_x-1) \times n_z) \tag{22}$$

对于等维反演勘探方案而言，其核矩阵 $G$ 由多个 $G_{t,n}$ 子矩阵组成的矩阵构成

$$G = \begin{pmatrix} G_{1,1} & G_{1,2} & \cdots & \cdots & G_{1,n} \\ G_{2,1} & G_{2,2} & \cdots & G_{2,o} & \vdots \\ \vdots & \vdots & G_{s,o} & \ddots & \vdots \\ \vdots & G_{s,2} & \ddots & \ddots & G_{t-1,n} \\ G_{t,1} & \cdots & \cdots & G_{t,n-1} & G_{t,n} \end{pmatrix}$$

## 3 模型试验
### 3.1 计算精度验证

以验证本文正演方法的有效性，特设置如下地球物理模型。在地下空间存在一个大小为 300 m ×300 m ×300 m、顶板埋深为500 m、底板为800 m且剩余密度为 $0.3 \times 10^3$ kg/m³ 的异常体。将异常体中心于地面的投影作为笛卡尔坐标系的原点，$n_x$、$n_y$ 和 $n_z$ 分别为40、40和30；

观测点间距为50 m，观测点个数为40 ×40=1600，观测点高度为地面上50 m，物性网格数个为40 ×40 ×20= 32000。

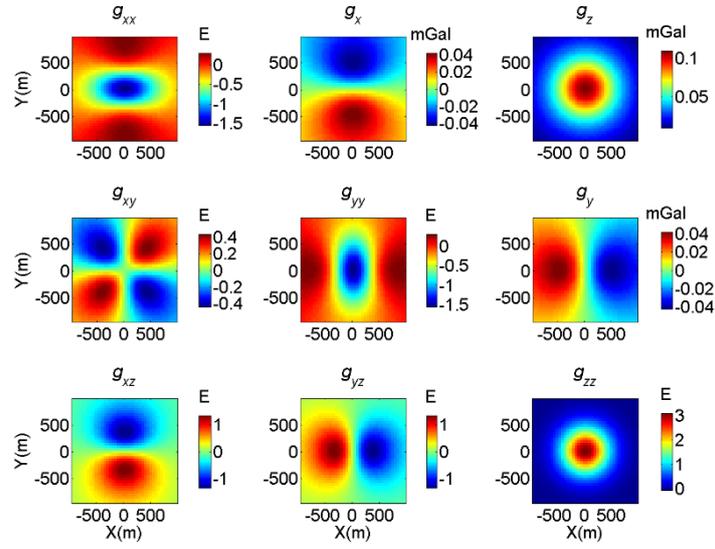

图 6　重力场一阶导数及重力梯度张量快速正演结果

Fig.6　Reuslts of forward the frist and second-order derivative of gravity based on rapid algrithm

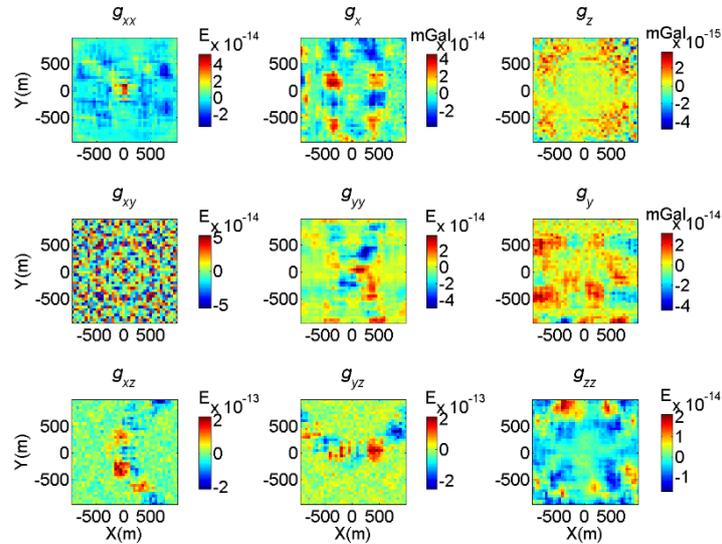

图 7　重力场一阶导数及重力梯度张量快速正演结果残差

Fig.7　Residual error of forward the frist and second-order derivative of gravity based on rapid algrithm

图 6和图 7分别为重力场一阶导数及重力梯度张量快速正演结果及其残差。通过以上两图的对比分析，可以发现正演计算残差比正演计算结果要小13个数量级。表明本文快速正演算法具有极高的精度。

### 3.2 内存消耗及计算性能对比分析

为对比以上三种算法的计算性能，设置一系列不同规模的地球物理模型。为使这些模型

不失代表性，设定模型中的物性网格的密度值是随机的以代表不同的复杂地质构造。为便于对比分析及刻画坐标轴，特设定观本组测试的观测网格为$n_x \times n_y = 32 \times 32$，进而对比不同模型规模重力$g_z$和重力梯度张量分量$g_{xx}$正演计算的耗时及内存占用。这里采用的硬件环境为：CPU配置为Intel T6570，主频2.10GHz，4G内存；软件环境为：Windows 7 64位操作系统，Matlab 2009a 开发编译环境。以下计算时间和用户内存消耗峰值为计算循环20次取平均所得的统计平均值。用户内存消耗峰值指在计算程序运行过程中正在使用的内存峰值。

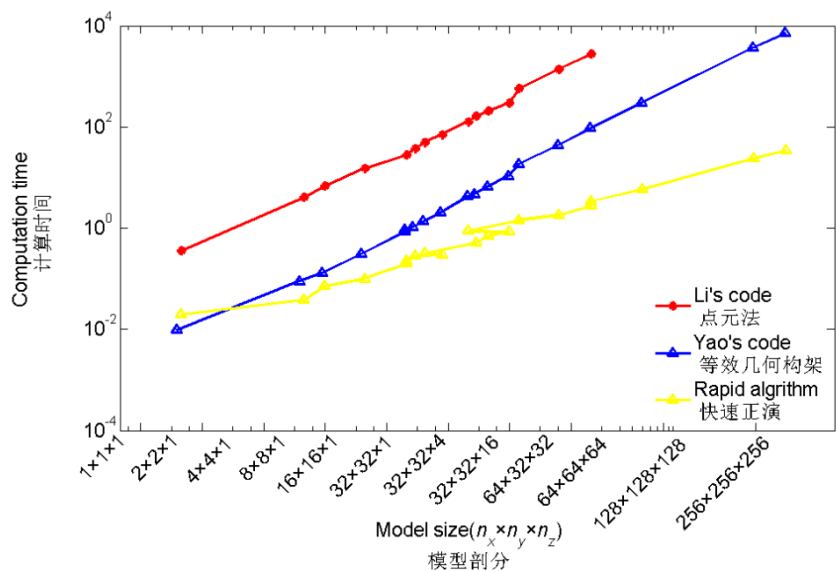

图 8　$g_z$ 正演模拟计算时间与模型规模关系曲线

Fig. 8　Comparing time-consuming of forward of $g_z$ component with different sizes models

从图 8中可以看出，点元法的计算时间最长，随着模型规模的增大，计算时间急剧增加；等效几何构架的计算时间比前者约小一个数量级。与前两类算法不同，当数据规模小于$32 \times 32 \times 1$时，此时处理的数据量很小，快速正演算法相对于等效几何构架的要慢，其主要时间消耗在构建式(19)中的系数$a$；而当数据规模大于$32 \times 32 \times 1$时，本研究所提出的算法的曲线更为平缓，表明其随模型规模增大，具有良好的可加速潜力。

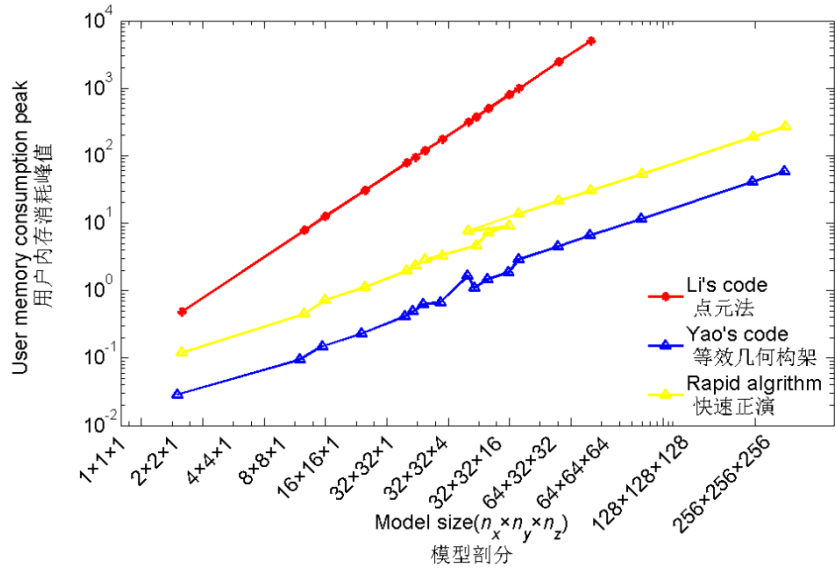

图 9  $g_z$ 正演模拟用户内存消耗峰值与模型规模曲线

Fig.9  Comparing peak memory usage of $g_z$ component with different sizes models

图 9为$g_z$正演模拟用户内存消耗峰值与模型规模曲线。当用户内存消耗峰值大于计算机内存时，计算设备需要扩充缓存或页面文件来确保程序正常运行，将极大的降低程序的计算效率。从图 9中三类正演方法的对比分析中可以看出，三类算法的用户内存消耗峰值与模型规模曲线均呈现很好的双对数线性关系。其中点元法的用户内存消耗峰值比等效几何构架的要大1~2个数量级，但这种关系并不随着模型规模的增大而变化；快速正演算法的用户内存消耗峰值最小。

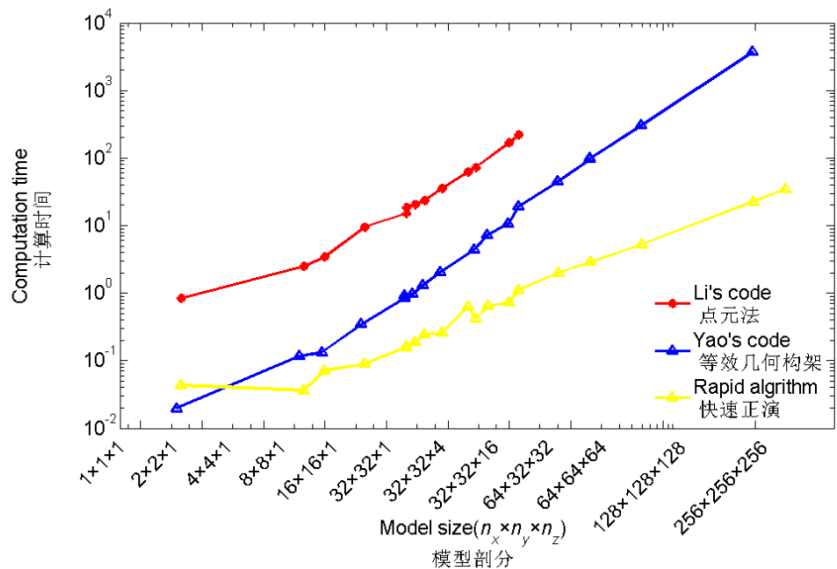

图 10  $g_{xx}$正演模拟计算时间与模型规模关系曲线

Fig. 10  Comparing time-consuming of forward of $g_{xx}$ component with different sizes models

通过图 8与图 10对比发现，使用三类模式计算$g_{xx}$的计算时间的关系与计算$g_z$的一致；

对于同一尺度的模型，$g_{xx}$的计算时间比$g_z$的要小，这与Li and Chouteau（1998）一文中对$g_{xx}$和$g_z$的计算复杂度描述一致；对于不同尺度的模型，随着模型规模的增加，快速正演算法的计算性能逐渐显现，这也验证了算法的可靠性。

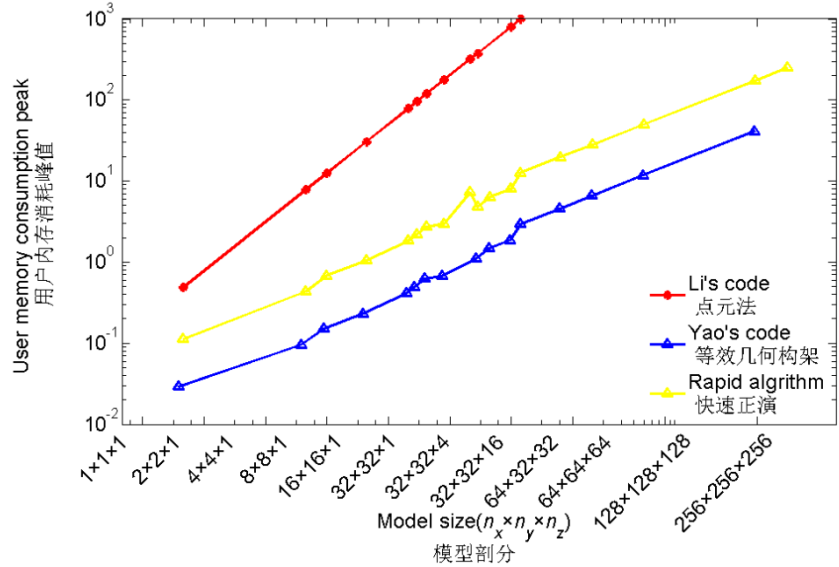

图 11  $g_{xx}$正演模拟用户内存消耗峰值与模型规模曲线

Fig. 11  Comparing peak memory usage of $g_{xx}$ component with different sizes models

通过图 9与图 11对比发现，以上三种算法的$g_{xx}$正演计算的用户内存峰值均与$g_z$的一致，也与式(5)、式(9)和式(22)式所描述的需要存储的核函数个数的关系一致。通过对比不同大小的模型，可以发现随着模型规模的不断增大其内存占用比率$Ratio_{G_{t,n}}$逐渐的减小，本文所提算法对内存的优化将更为明显。同时也表明对于相同大小的模型，本研究所提出的算法可以使用极小的内存而获得极大地加速潜力。

## 4  结论与建议

本文在等效几何构架技术的基础上，针对其应用于重力场正演计算时依赖于重力场的对称性，这导致其难于扩展至重力梯度正演计算。在针对重力/重力梯度各分量均具有平移等效性，本文作者引入与原有物性网格相同的三个虚拟网格将等效几何技术扩充至重力梯度正演技术。并进一步证明了重力/重力梯度各分量的核矩阵为BTTB矩阵，利用分块Toeplitz矩阵与向量相乘的快速计算方法，解决了重力/重力梯度数据反演解释的核矩阵的存储和运算问题。

1）  在基于等效几何构架的重力正演计算中，其依赖于重力场对称性。

2）  对于重力梯度张量所有分量而言，均具有平移等效性。基于此，引入虚拟网格构建虚拟物性网格，使得任意仅依赖于平移等效性计算，而不依赖于位场对称性。

3）  在重力及重力梯度张量正演计算中，相比于点元法和等效几何构架等两种算法，对于小尺度模型，基于本文所提算法的计算效率不高，而基于等效几何构架的正演计算相对较快；而对于大尺度模型，本文所提算法的计算具有极大的加速潜力。